\documentclass[a4paper]{report}
\usepackage[utf8]{inputenc}
\usepackage[T1]{fontenc}
\usepackage{RJournal}
\usepackage{amsmath,amssymb,array}
\usepackage{booktabs}
\usepackage{multicol}

\begin{document}

\sectionhead{Contributed research article}
\volume{XX}
\volnumber{YY}
\year{20ZZ}
\month{AAAA}

\begin{article}
  \title{logitFD: an R package for functional principal component logit regression}
  \author{by Manuel Escabias, Ana M. Aguilera and Christian Acal}
  
  \maketitle
  
  \abstract{
  	The functional logit regression model was proposed by \cite{Escabias04} with the objective of modeling a scalar binary response variable from a functional predictor. The model estimation proposed in that case was performed in a subspace of $L^2(T)$ of squared integrable functions of finite dimension, generated by a finite set of basis functions. For that estimation it was assumed that the curves of the functional predictor and the functional parameter of the model belonged to the same finite subspace. The estimation so obtained was affected by high multicollinearity problems and the solution given to these problems was based on different functional principal component analysis. The \CRANpkg{logitFD} package introduced here provides a toolbox for the fit of these models by implementing the different proposed solutions and by generalizing the model proposed in 2004 to the case of several functional and nonfunctional predictors. The performance of the functions is illustrated by using data sets of functional data included in the \CRANpkg{fda.usc} package from R-CRAN.
  }
  
  \section{Introduction}
  
  A functional variable is that whose values depend on a continuous magnitude such as time. They are functional in the sense that they can be evaluated at any time point of the domain, instead of the discrete way, in which they were originally measured or observed (see for example \cite{Ramsay05}). Different approaches have been used for the study of functional data, among others, the nonparametric methods proposed by \cite{Muller05} and \cite{FerratyVieu} or the basis expansion methods considered in \cite{Ramsay05}. Most multivariate statistical techniques have been extended for functional data, whose basic theory and inferential aspects are collected in recent books by \cite{Horvath}, \cite{Zhang2014} and \cite{Kokoszka2018}. The basic tools to reduce the dimension of the functional space to which the curves belong, are Functional Principal and Independent Component Analysis (FPCA) (\cite{Ramsay05}; \cite{Acal20}; \cite{Vidal2021}) and Functional Partial Least Squares (FPLS) (\cite{Preda2005}; \cite{Aguilera2010}; \cite{Aguilera2016}).

  In the last decade of the XXth century and the first decade of XXIth century, where functional data methods began to be developed, there was no adequate software available for using and fitting functional data methods. In fact, nowadays classical statistical software like SPSS, STATA,... do not have a toolbox for functional data analysis. The development of object-oriented software like R, Matlab or S-plus and the great activity of scientific community in this field has made possible to emerge different packages mainly in R for using functional data analysis (FDA) methods. Every package is designed from the point of view followed by its developer and the method used to fit functional data methods. For example \cite{Febrero2012} used nonparametric methods in their \CRANpkg{fda.usc} package, \cite{Ramsay09} designed their \CRANpkg{fda} package under basis expansion philosophy, Principal Analysis by Conditional Estimation (PACE) algorithm (see \cite{Zhu2014}) was used for curves alignment, PCA and regression in the \CRANpkg{fdasrvf} package (see \url{https://cran.r-project.org/web/packages/fdasrvf/index.html}). Recently Fabian Scheipl has summarized the available R packages for FDA (see \url{https://cran.r-project.org/web/views/FunctionalData.html}).
  
  This paper is devoted to \CRANpkg{logitFD} an R package for fitting the different functional principal component logit regression approaches proposed by \cite{Escabias04}. Functional logit regression is a functional method for modeling a scalar binary response variable in different situations: firstly, from one single functional variable as predictor; secondly, from several functional variables as predictors; and thirdly, from several functional and nonfunctional variables as predictors which is the most general case. There exist some R functions with this objective as the \code{fregre.glm} function of \CRANpkg{fda.usc} package (see \url{https://rpubs.com/moviedo/fda_usc_regression}). Different to the former the methods proposed by \cite{Escabias04}, and developed in \CRANpkg{logitFD}, are basis expansion based methods what makes the logit model suffer from multicollinearity. The proposed solutions were based on different functional principal components analysis: ordinary FPCA and filtered FPCA (see \cite{Escabias2014}). These models have been successfully applied to solve environmental problems (\cite{Aguilera20083187}; \cite{Escabias200595}; \cite{Escabias2013}) and classification problems in food industry (\cite{AguileraLisboa}). Extensions for the case of sparse and correlated data or generalized models have been also studied (\cite{James2002}; \cite{Muller05}; \cite{AguileraMorillo2012}; \cite{Mousavi2018}; \cite{Tapia2019}; \cite{Bianco2021}). 
  
  This package adopts \CRANpkg{fda}'s package philosophy of basis expansion methods of \cite{Ramsay09} and it is designed to use objects inherited from the ones defined in \CRANpkg{fda} package. For this reason \CRANpkg{fda} package is required for \CRANpkg{logitFD}. The package consists of four functions that fit a functional principal component logit regression model in four different situations
  \begin{itemize}
  	\item Filtered functional principal components of functional predictors, included in the model according to their variability explained power.
  	\item Filtered functional principal components  of functional predictors, included in the model automatically according to their prediction ability by stepwise methods.
  	\item Ordinary functional principal components of functional predictors, included in the model according to their variability explained power.
  	\item Ordinary functional principal components of functional predictors, included in the model automatically according to their prediction ability by stepwise methods.
  \end{itemize}
  
  The designed functions of our package use as input the \code{fd} objects given by \CRANpkg{fda} package and also provide as output \code{fd} objects among others elements.
  
  This paper is structured as follows: after this introduction, the second section shows the generalities of the package with the needed definitions and objects of functional data analysis, functional logit regression and extended functional logit regression, third and fourth sections board ordinary and filtered functional principal component logit regression, respectively. In fifth section ordinary and filtered functional principal components logit regression is addressed by including functional principal components according prediction ability by stepwise methods. In every section a summary of the theoretical aspects of the involved models is shown with a practical application with functional data contained in \CRANpkg{fda.usc} package (\cite{Febrero2012}).
  
  \section{\CRANpkg{logitFD} package: general statements}

  \subsection{Functional data analysis}
  
  A functional data set is a set of curves $ \left\{ x_1(t),\ldots, x_n (t) \right\}, $ with $t$ in a real interval $T$ ($ t \in T$). Each curve can be observed at different time points of its argument $t$ as $x_{i}=\left( x_{i}\left( t_{0}\right),\ldots ,x_{i}\left(t_{m_{i}}\right) \right)^{\prime}$ for the set of times $t_{0},\ldots,t_{m_{i}},\;i=1,\ldots ,n$ and these are not necessarily the same for each curve.
  
  Basis expansion methods assume that the curves belong to a finite dimensional space generated by a basis of functions $\left\{ \phi _{1}\left( t\right) ,\ldots ,\phi_{p}\left( t\right) \right\} $ and so they can be expressed as 
  \begin{equation}
  	x_{i}\left( t\right) =\sum_{j=1}^{p}a_{ij}\phi _{j}\left( t\right), \;
  	i=1,\ldots,n.
  	\label{BasisExpan}
  \end{equation}
  The functional form of the curves is determined when the basis coefficients $a_i=\left(a_{i1},\ldots,a_{ip}\right)^{\prime}$ are known. These can be obtained from the discrete observations either by least squares or by interpolation methods (see, for example, \cite{Escabias200595} and \cite{Escabias2006}).
  
  Depending on the characteristics of the curves and the observations, various types of basis can be used (see, for example, \cite{Ramsay05}). In practice, those most commonly used are, on the one had, the basis of trigonometric functions for regular, periodic, continuous and differentiable curves, and on the other hand, the basis of B-spline functions, which provides a better local behavior (see \cite{DeBoor2001}). In \CRANpkg{fda} package the type of basis used are B-spline basis, constant basis, exponential basis, Fourier basis, monomial basis, polygonal basis and power basis (\cite{Ramsay09}). Due to \CRANpkg{logitFD} package use \code{fd} objects from \CRANpkg{fda} package, the same types of basis can be used.
  
  In order to illustrate the use of \CRANpkg{logitFD} package we are going to use \code{aemet} data included in \CRANpkg{fda.usc} package of \cite{Febrero2012}. As can be read in the package manual, \code{aemet} data consist of meteorological data of 73 Spanish weather stations. This data set contains functional and nonfunctional variables observed in all the 73 weather stations. The information we are going to use to illustrate the use of our \CRANpkg{logitFD} package is the following:
  \begin{itemize}
  	\item \code{aemet\$temp}: matrix with 73 rows and 365 columns with the average daily temperature for the period 1980-2009 in Celsius degrees for each weather station.
  	\item \code{aemet\$logprec}: matrix with 73 rows and 365 columns with the average logarithm of precipitation for the period 1980-2009 for each weather station. We are going to use the proper precipitation, that is, \code{exp(aemet\$logprec)}
  	\item \code{aemet\$wind.speed}: matrix with 73 rows and 365 columns with the average wind speed for the period 1980-2009 for each weather station.
  	\item \code{aemet\$df[,c("ind","altitude","longitude","latitude")]}: data frame with 73 rows and 4 columns with the identifications code of each weather station, the altitude in meters over sea level and longitude and latitude of each weather station.
  \end{itemize}
  
  The problem with daily data is that they are too wiggly so if we need smooth curves with few basis functions, the loose of information is big. So, in order to illustrate the use of \CRANpkg{logitFD} package we are going to use mean monthly data. So for each one of the previously defined matrices we consider mean monthly data. On the other hand, \code{logprec} is also a very wiggly data set, so we considered their exponential. So the final data sets considered were the following:
  \begin{itemize}
  	\item \code{TempMonth}: matrix with 73 rows and 12 columns with the mean monthly temperature of \code{aemet\$temp}.
  	\item \code{PrecMonth}: matrix with 73 rows and 12 columns with the mean monthly exponential of \code{aemet\$logprec}.
  	\item \code{WindMonth}: matrix with 73 rows and 12 columns with the mean monthly wind speed of \code{aemet\$wind.speed}.
  \end{itemize}
  
  We are going to consider as binary response variable that variable with values: $1$ if a weather station is located in the north of Spain (above Madrid, the capital of Spain, and located in the geographic center of the country) and $0$ otherwise (stations of the south). Our objective will be to model the location of weather stations (north/south) from their meteorological information. This is a really artificial problem trying to explain the climate characteristics of Spanish weather stations classified according to their geographical location. Let us observe that only latitude is enough to determine the location of a weather station in the sense we are defining. In fact, latitude allows complete separation what makes the estimation of the logit model not to be possible (see for example \cite{Hosmer13}). 
  
  The steps for reading data would be
  
  \begin{verbatim}
  	library(fda.usc)
  	data(aemet)
  	Temp <- aemet$temp$data
  	Prec <- exp(aemet$logprec$data)
  	Wind <- aemet$wind.speed$data
  	StationsVars <- aemet$df[,c("ind","altitude","longitude","latitude")]
  	StationsVars$North <- c(1,1,1,1,0,0,0,0,1,1,0,0,0,0,0,1,1,1,0,0,1,0,0,0,0,1,0,0,1,1,1,1,1,
  	0,0,0,1,1,1,1,1,1,1,1,1,0,0,0,0,1,1,1,1,1,0,0,0,0,0,0,0,0,1,1,1,0,0,1,1,1,1,1,1)
  \end{verbatim}
  
  \noindent and the transformations to consider mean monthly data from daily data only for Temperature
  
  \begin{verbatim}
  	TempMonth <- matrix(0,73,12)
  	for (i in 1:nrow(TempMonth)){
  		TempMonth[i,1] <- mean(Temp[i,1:31])
  		TempMonth[i,2] <- mean(Temp[i,32:59])
  		TempMonth[i,3] <- mean(Temp[i,60:90])
  		TempMonth[i,4] <- mean(Temp[i,91:120])
  		TempMonth[i,5] <- mean(Temp[i,121:151])
  		TempMonth[i,6] <- mean(Temp[i,152:181])
  		TempMonth[i,7] <- mean(Temp[i,182:212])
  		TempMonth[i,8] <- mean(Temp[i,213:243])
  		TempMonth[i,9] <- mean(Temp[i,244:273])
  		TempMonth[i,10] <- mean(Temp[i,274:304])
  		TempMonth[i,11] <- mean(Temp[i,305:334])
  		TempMonth[i,12] <- mean(Temp[i,335:365])
  	}
  \end{verbatim}
  
  The rest of matrices (\code{PrecMonth} and \code{WindMonth}) were obtained in the same way.
  
  \CRANpkg{logitFD} is an R package for fitting functional principal component logit regression based on ordinary and filtered functional principal components described in previous sections. As was stated in the introduction, this package uses \CRANpkg{fda}'s package philosophy of basis expansion methods and it is designed to use objects inherited from the ones defined in \CRANpkg{fda} package. For this reason \CRANpkg{fda} package is required for \CRANpkg{logitFD}. The R functions designed in our package use as input the \code{fd} objects given by \CRANpkg{fda} package and also provide as output \code{fd} objects among others elements. In order to use our package it is assumed that the reader manage with \CRANpkg{fda} package, its objects and functions. 
  
  Let us begin with a brief explanation of the \CRANpkg{fda} objects required in our proposal. \CRANpkg{fda} package builds, from discrete observations of curves, an \code{fd} object (named \code{fdobj}) that will be used by \CRANpkg{logitFD} for its methods. So, let $X_{n\times m}=(x_i(t_k)),\; i=1,\ldots,n;\; k=1,\ldots,m$ be the matrix of discrete observations of curves $x_{1}\left( t\right) ,x_{2}\left( t\right) ,\ldots ,x_{n}\left( t\right) $ at the same time points $t_{1},t_{2},\ldots ,t_{m}$. An \code{fd} object is an \code{R} list with next elements:
  \begin{itemize}
  	\item \code{coefs}: the matrix of basis coefficients.
  	\item \code{basis}: an object of type \code{basis} with the information needed to build the functional form of curves based on basis expansion methods explained before. Depending on the selected basis the list of objects that contains the \code{basis} object can be different (see \CRANpkg{fda} reference manual).
  	\item \code{fdnames}: a list containing names for the arguments, function values and variables. This argument is not necessary.
  \end{itemize}
  The matrix of basis coefficients $A_{n \times p}=(a_{ij}), \; i=1,\ldots,n;\; j=1,\ldots,p$ (\code{coefs}) of all curves are obtained by least squares as $A^{T}=\left( \Phi ^{T}\Phi \right) ^{-1}\Phi ^{T}X^{T}$
  where  $\Phi_{m \times p} = (\phi _{j}\left( t_{k}\right)),\; j=1,\ldots,p; \; k=1,\ldots,m$ is the matrix of basis functions evaluated at sampling points.
  
  The \code{basis} object allows the basis expansion (\ref{BasisExpan}) of curves. We consider for aemet data these two basis:
  \begin{itemize}
  	\item $7$-length Fourier basis for Temperature.
  	\item $8$-length cubic B-spline basis for Precipitation and Wind
  \end{itemize}
  The \code{R} parameters needed to define the basis object depend on the type of basis used (see fda R reference manual). Fourier basis only needs the interval where basis functions are defined and the dimension of the basis. B-spline basis needs also the degree of polynomials that define the basis functions. The default degree is 3.
  
  The sentences to create the used basis have been
  
  \begin{verbatim}
  	FourierBasis <- create.fourier.basis(rangeval = c(1,12),nbasis=7)
  	BsplineBasis <- create.bspline.basis(rangeval = c(1,12),nbasis=8)
  \end{verbatim}
  
  The main function of \CRANpkg{fda} package that provides the \code{fdobj} object from discrete data in a matrix is \code{Data2fd} function (see \CRANpkg{fda} reference manual). Our \code{fdobj} were obtained with the sentences
  
  \begin{verbatim}
  	TempMonth.fd <- Data2fd(argvals = c(1:12), y=t(TempMonth),basisobj = FourierBasis)
  	PrecMonth.fd <- Data2fd(argvals = c(1:12), y=t(PrecMonth),basisobj = BsplineBasis)
  	WindMonth.fd <- Data2fd(argvals = c(1:12), y=t(WindMonth),basisobj = BsplineBasis)
  \end{verbatim}
  
\begin{figure}
  	\begin{center}
  		\begin{tabular}{ccc}
  			\includegraphics[width=0.33\textwidth]{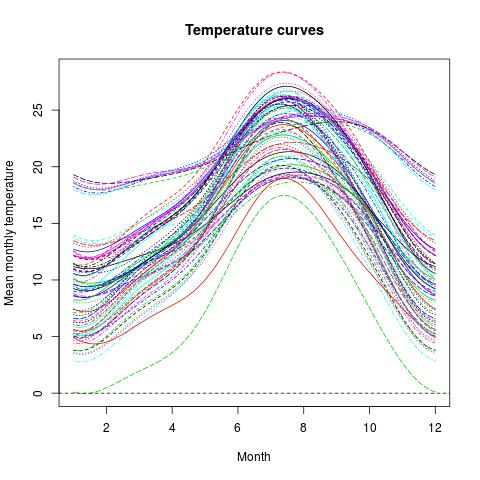} &  \includegraphics[width=0.33\textwidth]{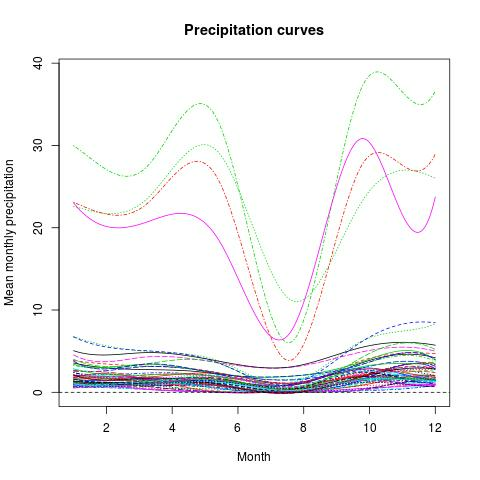} & \includegraphics[width=0.33\textwidth]{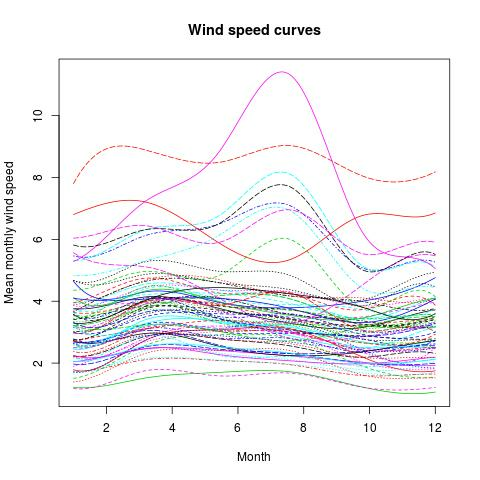} \\
  		\end{tabular}
  	\end{center}
  	\caption{Curves of mean monthly Temperature, Precipitation and Wind.}
  	\label{FDCurves}
  \end{figure}

  An \code{fdobj} allows plotting all curves by using the \code{R} \code{plot()} sentence. The functional data so obtained can be seen in Figure \ref{FDCurves}.
  
  \subsection{Functional logit regression model}
  
  In order to understand how the functions of the \CRANpkg{logitFD} work, let summarize the theoretical aspects of the models involved.
  
  Let $Y$ be a binary response random variable and let $\left\{ X_1\left( t\right),X_2\left( t\right),\ldots,X_R\left( t\right) :t\in T\right\} $ be a set of functional covariates related to $Y.$ Let $x_{11}\left( t\right) ,\ldots ,x_{n1}\left( t\right),\ldots,x_{1R}\left( t\right) ,\ldots ,x_{nR}\left( t\right) $ be $R$ samples of curves of the functional predictors that can be summarized by columns in a matrix of curves
  \begin{equation}
  	\left( 
  	\begin{array}{cccc}
  		x_{11}\left( t\right) & x_{12}\left( t\right) & \cdots & x_{1R}\left(
  		t\right) \\ 
  		x_{21}\left( t\right) & x_{22}\left( t\right) & \cdots & x_{2R}\left(
  		t\right) \\ 
  		\cdots & \cdots & \cdots & \cdots \\ 
  		x_{n1}\left( t\right) & x_{n2}\left( t\right) & \cdots & x_{nR}\left(
  		t\right)
  	\end{array}
  	\right) \label{PredictorsMatrix}
  \end{equation}
  Let $y_{1},\ldots,y_{n}$ be a sample of the binary response associated with the curves ($y_i \in \{0,1\}$), then the functional logit model in terms of the functional predictors is formulated as
  \begin{equation}
  	y_{i}=\pi _{i}+\varepsilon _{i}=\pi \left( x_{i1}\left( t\right),\ldots,x_{iR}\left( t\right) \right)
  	+\varepsilon _{i} \Leftrightarrow \pi _{i}=\dfrac{\exp \left\{ l_i \right\} }{1+\exp \left\{ l_i\right\} },\;i=1,\ldots ,n,  \label{probfun}
  \end{equation}
  where $\boldsymbol{\varepsilon} =\left( \varepsilon _{1},\ldots
  ,\varepsilon _{n}\right)
  ^{\prime}$ is the vector of independent centered random errors, with unequal variances and Bernoulli distribution, and $l_i$ (known as logit transformations) are modelized from functional predictors as
  \begin{eqnarray}
  	l_{i}&=&\ln \left[ \dfrac{\pi _{i}}{1-\pi _{i}}\right] =\alpha
  	+\int_{T}x_{i1}\left( t\right) \beta_1 \left( t\right) dt+\int_{T}x_{i2}\left( t\right) \beta_2 \left( t\right) dt+\cdots+\int_{T}x_{iR}\left( t\right) \beta_R \left( t\right) dt.
  \end{eqnarray}
  This model has $R$ functional parameters to be estimated $\beta_1 \left( t\right),\ldots,\beta_R \left( t\right).$ If we consider that the curves of each functional predictor belong to a finite space generated by a basis of functions as in (\ref{BasisExpan}) and that the corresponding functional parameter belongs to the same space (same basis for each pair $(X_r(t),\beta_r(t)), \; r=1,\ldots,R$)
  \begin{eqnarray}
  	\beta_r\left( t\right) =\sum_{l=1}^{p_r }\beta_{rl} \phi _{rl}\left( t\right), \; r=1,\ldots,R,
  	\label{ParamFunBasis}
  \end{eqnarray}
  the functional logit model in terms of the logit transformations is expressed in matrix form as
  \begin{eqnarray}
  	L&=& \mathbf{1}\alpha + A_1 \Psi_1 \beta_1 + A_2 \Psi_2 \beta_2 + \cdots + A_R \Psi_R \beta_R,
  \end{eqnarray}
  where
  \begin{itemize}
  	\item  $L=\left( l_{1},\ldots ,l_{n}\right) ^{\prime }$ is the vector of logit
  	transformations. 
  	
  	\item  $\left( \mathbf{1}\;|\;A_1\Psi_1\;|\;A_2\Psi_2\;| \cdots |\;A_R\Psi_R \right) $ is the design matrix, and $|$
  	indicating the separation between the boxes of the matrix.
  	
  	\item  $\mathbf{1}=\left( 1,\ldots ,1\right) ^{\prime }$ is a $n-$length vector
  	of ones.
  	
  	\item  $\Psi_r=(\psi_{jkr}),\; r=1,\ldots,R $ is the matrix whose entries are the inner products between basis functions of the space where curves belong to
  	\begin{eqnarray}
  		\psi_{jkr}=<\phi_{jr}(t),\phi_{kr}(t)>=\int_T \phi_{jr}(t)\phi_{kr}(t)dt, \; j,k=1,\ldots,p_r; \; r=1,\ldots,R.
  		\label{inprod}
  	\end{eqnarray}
  	
  	\item  $A_r,\; r=1,\ldots,R$ is the matrix of basis coefficients as rows of sample curves of the space where curves belong to.
  	
  	\item  $\beta_r =\left( \beta_{r1},\ldots ,\beta_{rp_r}\right)^{\prime },\; r=1,\ldots,R$ are the basis coefficients of the
  	functional parameter $\beta_r(t),\; r=1,\ldots,R$.
  	
  \end{itemize}
  
  Let us observe that each functional predictor (and functional parameter) can be expressed in terms of a different type of basis and different number of basis functions.
  
  This functional logit model provides severe multicollinearity problems as was stated in \cite{Escabias04} for the case of a single functional predictor that was the original formulation of the model.
  
  \subsection{Extended functional logit model: several functional and nonfunctional predictors}
  
  We can finally formulate the functional logit model in terms of more than one functional predictor and non-functional ones. So let $Y$ be a binary response variable and let $\left\{X_1\left( t\right),X_2\left( t\right),\ldots,X_R\left( t\right): t\in T\right\} $ be a set of functional covariates related to $Y$ and $U_1,U_2,\ldots,U_S$ a set of non-functional predictors. Let us consider the sample of curves (\ref{PredictorsMatrix}) and
  \begin{equation*}
  	\left( 
  	\begin{array}{cccc}
  		u_{11} & u_{12} & \cdots & u_{1S} \\ 
  		u_{21} & u_{22} & \cdots & u_{2S} \\ 
  		\cdots & \cdots & \cdots & \cdots \\ 
  		u_{n1} & u_{n2} & \cdots & u_{nS}
  	\end{array}
  	\right)
  \end{equation*}
  the sample of observations of nonfunctional predictors, let $y_{1},\ldots,y_{n}$ be a sample of the response associated with the curves. Then the model is expressed in terms of logit transformations as
  \begin{eqnarray}
  	l_i =\alpha
  	+\int_{T}x_{i1}\left( t\right) \beta_1 \left( t\right) dt+\cdots+\int_{T}x_{iR}\left( t\right) \beta_R \left( t\right) dt + u_{i1} \delta_1+\cdots+u_{iS} \delta_S,\; i=1,\ldots ,n.
  	\label{pclogitfun2}
  \end{eqnarray}
  Now the model has $R$ functional parameters to estimate $\beta_1\left( t\right),\ldots,\beta_R \left( t\right)$ and $S$ nonfunctional parameters $\delta_1,\ldots,\delta_S.$ As in the previous case, each functional predictor and functional parameter can be expressed in terms of a different type of basis and different number of basis functions as in (\ref{BasisExpan}) and (\ref{ParamFunBasis}). We consider again the same basis for each pair $(X_r(t),\beta_r(t)), \; r=1,\ldots,R.$ The functional logit model in terms of the logit transformations is expressed in matrix form as
  \begin{eqnarray*}
  	L&=& \mathbf{1}\alpha + A_1 \Psi_1 \beta_1 + A_2 \Psi_2 \beta_2 + \cdots + A_R \Psi_R \beta_R+U_1 \delta_1+\cdots+U_S \delta_S.
  \end{eqnarray*}
  This model has as only difference with respect the previous one the design matrix of the model $\left( \mathbf{1}\;|\;A_1\Psi_1\;|\;A_2\Psi_2\;| \cdots |\;A_R\Psi_R \;| U_1 | \cdots | U_S \right), $ where $U_1,\ldots,U_S$ represent the columns of observations of the nonfunctional predictors, and a set of scalar parameters $\delta_1,\ldots,\delta_S.$ As in the previous case, this model has multicollinearity problems.
  

  \section{Ordinary functional principal components logit regression}
  
  The proposed solution to solve the multicollinearity problems in \cite{Escabias04} for the single model (only one functional predictor) was to use as predictors a set of functional principal components. Let us briefly remember the functional principal component analysis principles.
  
  Let $x_{1}\left( t\right) ,\ldots ,x_{n}\left( t\right) $ be a set of curves with mean curve and covariance surface respectively
  \begin{equation*}
  	\overline{x}\left( t\right) =\dfrac{1}{n}\sum_{i=1}^{n}x_{i}\left( t\right), \; C\left( s,t\right) =\dfrac{1}{n-1}\sum_{i=1}^{n}\left( x_{i}\left(
  	s\right) -\overline{x}\left( s\right) \right) \left( x_{i}\left( t\right) -%
  	\overline{x}\left( t\right) \right).
  \end{equation*}
  Functional principal components are defined as
  \begin{equation*}
  	\xi _{ij}=\int_{T}\left(x_{i}\left( t\right)-\overline{x}\left( t\right) \right) f_j\left( t\right) dt, \; f_{j}\left( t\right) =\sum_{k=1}^{p}F_{jk}\phi _{k}\left( t\right), \;  j=1,\ldots,p; \; i=1,\ldots,n.
  \end{equation*}
  In this formulation it is assumed that curves are expressed as in (\ref{BasisExpan}), and, as a consequence, the eigenfunctions $f_j(t),$ that define the functional principal components, are also basis expansion expressed, being the basis coefficients $F_j$ the eigenvectors of $A\Psi^{1/2}$ matrix (see \cite{Ocana2007}). For a more general and detailed situation see \cite{Ramsay05}. The original curves can be expressed in terms of the functional principal components as
  \begin{equation*}
  	x_{i}\left( t\right) =\overline{x}(t)+\sum_{j=1}^{p }\xi _{ij}f_{j}\left( t\right)=\overline{x}(t)+\sum_{j=1}^{p } \sum_{k=1}^{p} \xi _{ij} F_{jk}\phi _{k}\left( t\right),\;i=1,\ldots ,n.
  \end{equation*}
  If a reduced set of functional principal components is considered, the original curves can be approximated by
  \begin{equation}
  	x_{i}\left( t\right) \simeq \overline{x}(t) + \sum_{j=1}^{q<p }\xi _{ij}f_{j}\left( t\right)=\overline{x}(t)+\sum_{j=1}^{q<p } \sum_{k=1}^{p} \xi _{ij} F_{jk}\phi _{k}\left( t\right),\;i=1,\ldots ,n \label{FPCA}
  \end{equation}
  The quality of this approximation will depend on the percentage of explained variability that acumulates the first $q$ functional principal components, given by 
  \begin{equation*}
  	\dfrac{\sum_{j=1}^q \lambda_j}{\sum_{j=1}^p \lambda_j}.
  \end{equation*}
  
  The ordinary functional principal components logit regression solution to solve the multicollinearity problems of the functional logistic regression model consists of considering a functional principal component expansion of each sample curve for each functional predictor and turning the functional model into a multivariate one whose covariates are the considered functional principal components. The number of principal components required can be different in each functional predictor, but the same for all curves of a specific functional predictor.
  
  In order to get an estimation of the functional parameter for the case of a single functional covariate, by considering the principal component expansion of curves, the logit model adopts the following expression
  \begin{eqnarray*}
  	l_{i}&=&\alpha +\int_{T} \left(\overline{x}(t) + \sum_{j=1}^{p }\xi _{ij}f_{j}\left( t\right) \right)  \beta \left( t\right)dt
  	=\alpha
  	+\int_{T} \overline{x}(t) \beta \left( t\right)dt + \sum_{j=1}^{p }\xi _{ij} \int_{T} f_{j}\left( t\right) \beta \left( t\right)dt \\
  	&=&\gamma_0+\sum_{j=1}^{p }\xi _{ij} \gamma_j,\;i=1,\ldots ,n.
  \end{eqnarray*}
  These expressions enables to express the basis coefficients of the functional parameter and the intercept parameter of the logit model in terms of the parameters estimated from the functional principal components of the curves.
  \begin{eqnarray}
  	\alpha &=& \gamma_0 - \int_{T} \overline{x}(t) \beta \left( t\right)dt = \alpha +(\overline{a}_1,\ldots,\overline{a}_p) \Psi (\beta_1,\ldots\beta_p)^{\prime} \label{Intercept} \\
  	(\beta_1,\ldots,\beta_p)^{\prime} &=& \Psi F (\gamma_1,\ldots, \gamma_p)^{\prime}
  	\label{pclogitfun}
  \end{eqnarray}
  with $\Psi=\left(\psi _{jk}\right)$ being the inner products between the basis functions (as in (\ref{inprod})) and $F$ the orthogonal matrix of basis coefficients of principal component curves shown in (\ref{FPCA}).
  
  If we consider the principal component expansion of curves in terms of a reduced set of functional principal components we can get an estimation of the basis coefficients of the functional parameter whose accuracy depends on the accumulated variability of the selected principal components (see \cite{Escabias04}).
  
  So, if we denote by $\Gamma_1,\Gamma_2,\ldots,\Gamma_R$ the ordinary functional principal components matrices of the sample curves associated with the functional predictors $\left\{ X_1\left( t\right),X_2\left( t\right),\ldots,X_R\left( t\right) :t\in T\right\}, $ respectively, the functional principal component logit model in terms of the logit transformations is expressed in matrix form as
  \begin{eqnarray*}
  	L&=& \mathbf{1}\alpha + \Gamma_1 \gamma_1 + \Gamma_2 \gamma_2 + \cdots + \Gamma_R \gamma_R+U_1 \delta_1+\cdots+U_S \delta_S,
  \end{eqnarray*}
  where $\left( \mathbf{1}\;|\;\Gamma_1\;|\;\Gamma_2\;| \cdots |\;\Gamma_R \right|U_1 | \cdots |U_S ) $ is the design matrix in terms of ordinary functional principal components, $\gamma_r =\left( \gamma_{r1},\ldots ,\gamma_{rp_r}\right)^{\prime }$ are the coefficients of the multiple model associated to the corresponding functional principal components and $\left( \delta_{1},\ldots ,\delta_{s}\right)^{\prime }$ the scalar parameters associated to non-functional variables. By using a reduced set of $q_1,q_2,\ldots,q_R$ functional principal components, being the scores matrix denoted as $\Gamma_{1(q_1)},\Gamma_{2(q_2)},\ldots,\Gamma_{R(q_R)},$ respectively, the model is then expressed as
  \begin{eqnarray*}
  	L&=& \mathbf{1}\alpha + \Gamma_{1(q_1)} \gamma_1 + \Gamma_{2(q_2)} \gamma_2 + \cdots + \Gamma_{R(q_R)} \gamma_R +U_1 \delta_1+\cdots+U_S \delta_S.
  \end{eqnarray*}
  
  Basis coefficients for each functional parameter are then obtained by formula (\ref{pclogitfun}) from their corresponding $\gamma$ parameter and the intercept $\alpha$ by formula (\ref{Intercept}).
  
  \code{logitFD.pc} is the function from \CRANpkg{logitFD} package that fits the ordinary functional principal component logit regression model. The declaration of the function has this form: \begin{center} \code{logitFD.pc(Response, FDobj=list(), ncomp=c(), nonFDvars=NULL)}, \end{center} and the function arguments are the following:
  
  \begin{itemize} 
  	\item \code{Response}: vector of responses $y_{1},\ldots ,y_{n}.$ 
  	\item \code{FDobj}: list of the different functional objects (\code{fdobj}) to use from the \code{fd} package. Theoretically $x_1(t),\ldots,x_R(t).$
  	\item \code{ncomp}: vector with the number of functional principal components $q$ to use in the model for each functional predictor. The length of the vector must be equal to the length of the \code{FDobj} list. The first element of the vector corresponds with the number of functional principal components of the first functional predictor (columns of $\Gamma_1$), the second with the columns of $\Gamma_2$, $\ldots$, the $R$th with the columns of $\Gamma_R.$
  	\item \code{nonFDvars}: data frame with the observations of the scalar predictor variables, that is, with columns $U_1\ldots,U_S.$ Let us observe that the number of rows of this data frame must be the same as the length of the response vector. Likewise, the number of functions in each functional object must be the same for all functional objects.
  \end{itemize} 
  
  In order to illustrate the performance of the function, let us consider \code{StationsVar\$North} as binary response variable, \code{TempMonth} and \code{PrecMonth} as functional predictors, and as scalar predictor variables \code{StationsVar[,c("altitude","longitude")]}. We are going to consider the first 3 and 4 functional principal components of \code{TempMonth} and \code{PrecMonth} respectively.
  
  Our fit is obtained as
  
  \begin{verbatim}
  	Fit1 <- logitFD.pc(Response=StationsVars$North,FDobj=list(TempMonth.fd,PrecMonth.fd),
  	ncomp = c(3,4),nonFDvars = StationsVars[,c("altitude","longitude")])
  \end{verbatim}
  
  The output of the function is an \code{R} list with objects: \code{glm.fit}, \code{Intercept}, \code{betalist}, \code{PC.variance} and \code{ROC.curve}. These elements are explained next.
  
  \paragraph{\code{glm.fit} object of \code{Fit1}:} Object of class inherited from \code{"glm"}. This object contains details about the fit of the multiple logit model to explain the binary response from the selected functional principal components and the scalar variables. This output allows to use different R functions as \code{summary()} function to obtain or print a summary of the fit, or \code{anova()} function to produce an analysis of variance table, and to extract various useful features of the values returned by \code{"glm"} as {\it coefficients}, {\it effects}, {\it fitted.values} or {\it residuals} (see R help). In our example the summary of the fit can be seen on page (\pageref{glmsummary}). Let us observe that the package assigns the names \code{A.1}, \code{A.2} and \code{A.3} and \code{B.1}, \code{B.2}, \code{B.3} and \code{B.3} to the first 3 and 4 functional principal components of the functional covariates. From this object it would easily be able to make an analysis of residuals, with \code{residuals()} function, or fitted values, with \code{fitted.values()} function, testing goodness of fit, etc. A classical goodness of fit measure is the correct classification rate (CCR) from the classification table. In our example both elements can be easily obtained through these sentences \code{table(StationsVars\$North,round(predict(Fit1\$glm.fit, type = "response")))} and \code{100*sum(diag(table(StationsVars\$North, round(predict(Fit1\$glm.fit, type = "response")))))/nrow(StationsVars)}. From the results we can conclude that if we want to model the weather stations location from the temporal evolution of temperatures and precipitation and from altitude and longitude variables, we classify correctly 94.5\% of stations.
  
  \paragraph{\code{Intercept}  object of \code{Fit1}:} The $\alpha$ (intercept) estimated parameter through expression (\ref{Intercept}) is given in the object \code{Fit1\$Intercept}.
  
\begin{figure}
  	\begin{center}
  		\begin{tabular}{ccc}
  			\includegraphics[width=0.33\textwidth]{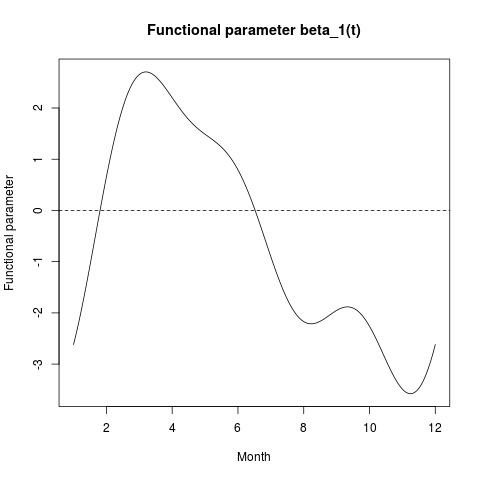} & \includegraphics[width=0.33\textwidth]{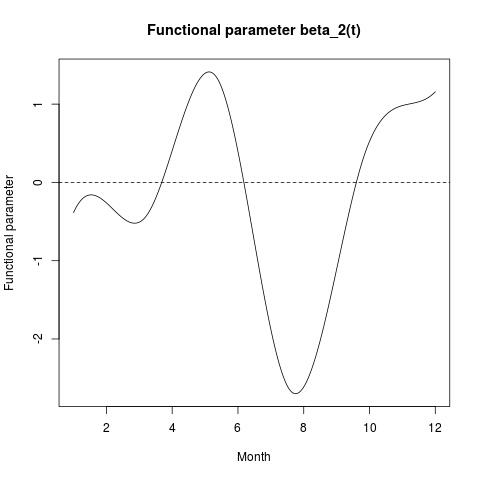} &
  			\includegraphics[width=0.33\textwidth]{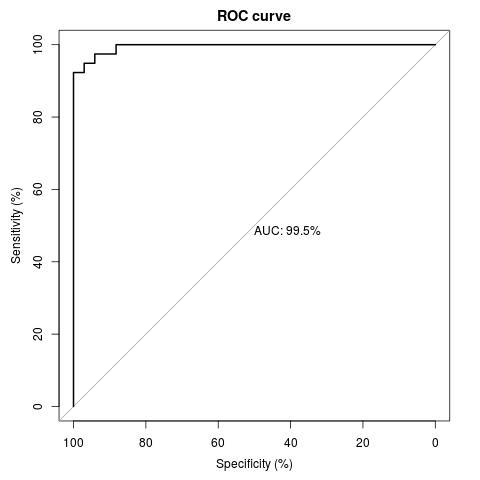} \\
  			$\hat{\beta}_1(t)$ & $\hat{\beta}_2(t)$ & ROC curve \\
  		\end{tabular}
  	\end{center}
  	\caption{Functional parameters and Roc Curve from Fit 1 by means of \code{logitFD.pc()} function.}
  	\label{Fit1}
  \end{figure}

  \paragraph{\code{betalist} object of \code{Fit1}:} list of functional objects. Each element of the list contains the functional parameter corresponding to the associated functional predictor variable located in the same position of \code{FDobj} parameter that appears in the function. In our case, firstly temperature curves where introduced, whereas precipitation curves were added in second place. Then the first two elements of \code{betalist}, that is, \code{[[1]]} and \code{[[2]]} will be the functional parameters associated with temperature and precipitation curves respectively. If we use more functional data, \code{[[3]], [[4]],...} provide the corresponding functional parameters. Let us remember that as \code{fdobj}, its elements are \code{coefs}: the matrix (vector in this case) of basis coefficients, \code{basis}: the same basis used in \code{FDobj} object and the rest elements as \code{fdnames}. Besides, multiple functions from \code{fd} package can be used such as the \code{plot()} function, used here as \code{plot(Fit1\$betalist[[1]])} and \code{plot(Fit1\$betalist[[2]])} for the parameter  functions associated to Temperature and Precipitation curves respectively. The plots that generate these sentences can be seen in Figure \ref{Fit1}. We could also evaluate these functions in a grid with the function \code{eval.fd()}, for example in the observed months-time, we could obtain the values on page (\pageref{BasisEval}).
  
  \paragraph{\code{PC.variance} object of \code{Fit1}:} list of {\it data.frames} with explained variability of functional principal components. Each element of the list contains the cumulative variance matrix corresponding to the associated functional variable in the same position. In our case, the first input curves were temperature curves and the second ones, the precipitation curves. In this point, the first element \code{[[1]]} of \code{PC.variance} will be the matrix of explained variability of functional principal components associated with temperature curves whereas the second element \code{[[2]]} of the \code{PC.variance} will be the matrix of explained variability of principal components associated with precipitation curves as with \code{betalist}. If we use more functional data \code{[[3]], [[4]],...} the function provides the corresponding explained variability matrices. \noindent The output got in \code{PC.variance} list is on page (\pageref{VarAcum}). We can observe that the first two functional principal component of temperature and precipitation accumulate 99.4\% and 99.1\% of the total variability respectively, so that the selection of 3 components for temperature and 4 for precipitation are enough for a good representation of the curves.
  
  \paragraph{\code{ROC.curve}  object of \code{Fit1}:} an object of the \code{roc()} function from \CRANpkg{pROC} package whose mission is to test the prediction ability of the model. This function builds a ROC curve and returns a \code{roc} object, i.e. a list of class \code{roc}. This object can be printed, plotted, or passed to many other functions (see reference manual). As default this element returns the area under the ROC curve with the object \code{Fit1\$ROC}. The plot of the ROC curve with sentence \code{plot(Fit1\$ROC)} can be seen in Figure \ref{Fit1}. As it was stated from the correct classification rate, the ROC curve and its graph allows us to observe that the fit is accurate for this modeling.
  
  \section{Filtered functional principal components included in the model according to their explained variability}
  
  Alternatively to ordinary functional principal component logit regression, \cite{Escabias2014} discussed a different approach based on equivalences proved by \cite{Ocana2007} and \cite{Ocana1999} between different functional principal component analysis. These equivalences stated that given $x_{1}\left( t\right) ,\ldots ,x_{n}\left( t\right) $ a set of curves, the functional principal component analysis of the transformed curves $L(x_1(t)),\ldots,L(x_n(t))$ defined by $L(x_i(t))=\sum_{j=1}^p a_{ij}^{\ast} \phi_j(t),$ being
  $(a_{i1}^{\ast},\ldots,a_{ip}^{\ast})^{\prime}=\Psi^{1/2} (a_{i1},\ldots,a_{ip})^{\prime},$ is equivalent to multivariate PCA of the design matrix $A\Psi$ associated with the functional logit model. In this expansion, the principal component curves $f_{j}^{\ast}\left( t\right)$ are expressed in terms of the basis functions as
  \begin{equation*}
  	f_{j}^{\ast}\left( t\right) =\sum_{k=1}^{p}F_{jk}^{\ast}\phi _{k}\left( t\right), \;
  	j=1,\ldots,p,
  \end{equation*}
  where basis coefficients in matrix form are obtained as $F^{\ast}=\Psi^{-1/2}V,$ being $V$ the eigenvectors of the covariance matrix of $A\Psi.$
  
  The original curves can be also approximated
  \begin{equation*}
  	L(x_{i}\left( t\right)) =L(\overline{x}(t))+\sum_{j=1}^{p }\xi_{ij}^{\ast}f_{j}^{\ast}\left( t\right),\;i=1,\ldots ,n,
  \end{equation*}
  where $\xi_{ij}^{\ast}$ are the functional principal components scores of the transformed curves $L(x_1(t)),\ldots,L(x_n(t)).$
  
  Now again the original curves can be approximated by using a reduced set of these functional principal components
  \begin{equation*}
  	L(x_{i}\left( t\right)) \simeq L(\overline{x}(t)) + \sum_{j=1}^{q<p }\xi_{ij}^{\ast}f_{j}^{\ast}\left( t\right),\;i=1,\ldots ,n.
  \end{equation*}
  
  In order to avoid multicollinearity in functional logit model an alternative is to use filtered principal components (see \cite{Escabias04}). So let $x_{1}\left( t\right) ,\ldots ,x_{n}\left( t\right) $ be a set of curves with mean curve $\overline{x}\left( t\right)$ and $y_{1},\ldots,y_{n}$ the response associated observations. Let $\Gamma^{\ast}=(\xi_{ij}^{\ast})$ be the $n \times p$ matrix of functional principal components, and $f_1^{\ast}(t),\ldots,f_p^{\ast}(t)$ the principal component curves. The filtered functional principal component logit regression can be expressed
  \begin{eqnarray*}
  	l_{i}&=&\alpha +\int_{T} \left(\overline{x}(t) + \sum_{j=1}^{p }\xi_{ij}^{\ast}f_{j}^{\ast}\left( t\right) \right)  \beta \left( t\right)dt
  	=\alpha
  	+\int_{T} \overline{x}(t) \beta \left( t\right)dt + \sum_{j=1}^{p }\xi_{ij}^{\ast} \int_{T} f_{j}^{\ast}\left( t\right) \beta \left( t\right)dt \\
  	&=&\gamma_0^{\ast}+\sum_{j=1}^{p }\xi_{ij}^{\ast} \gamma_j^{\ast},\;i=1,\ldots ,n.
  \end{eqnarray*}
  This expression also allows expressing the basis coefficients of the functional parameter and the intercept parameter of the logit model alternatively in terms of the parameters estimated from the filtered functional principal components of the curves equivalently to (\ref{pclogitfun}) and (\ref{Intercept}) respectively:
  \begin{eqnarray}
  	\alpha &=& \gamma_0^{\ast} - \int_{T} \overline{x}(t) \beta \left( t\right)dt = \alpha +(\overline{a}_1,\ldots,\overline{a}_p) \Psi (\beta_1,\ldots\beta_p)^{\prime} \\
  	(\beta_1,\ldots,\beta_p)^{\prime} &=& \Psi F^{\ast} (\gamma_1^{\ast},\ldots, \gamma_p^{\ast})^{\prime}
  	\label{filteredPCA}
  \end{eqnarray}
  due to $F^{\ast}$ and $\Psi$ matrices are orthogonal and $\Psi$ is also a symmetric matrix.
  
  If we consider a principal component expansion of curves in terms of a reduced set of filtered functional principal components we can get an estimation of the basis coefficients of the functional parameter whose accuracy depends of the accumulated variability of the selected principal components (see \cite{Escabias04}).
  
  So, if we denote by $\Gamma_1^{\ast},\Gamma_2^{\ast},\ldots,\Gamma_R^{\ast}$ the matrices of filtered functional principal components of the sample curves of the functional predictors $\left\{ X_1\left( t\right),X_2\left( t\right),\ldots,X_R\left( t\right) :t\in T\right\} $ respectively, the functional principal component logit model in terms of the logit transformations is expressed in matrix form as
  \begin{eqnarray*}
  	L&=& \mathbf{1}\alpha + \Gamma_1^{\ast} \gamma_1^{\ast} + \Gamma_2^{\ast} \gamma_2^{\ast} + \cdots + \Gamma_R^{\ast} \gamma_R^{\ast}+U_1 \delta_1+\cdots+U_S \delta_S,
  \end{eqnarray*}
  where $\left( \mathbf{1}\;|\;\Gamma_1^{\ast}\;|\;\Gamma_2^{\ast}\;| \cdots |\;\Gamma_R^{\ast} \right|U_1 | \cdots |U_S ) $ is the design matrix in terms of ordinary functional principal components, $\gamma_r^{\ast} =\left( \gamma_{r1}^{\ast},\ldots ,\gamma_{rp_r}^{\ast}\right)^{\prime }$ are the coefficients of the multiple model associated to the corresponding filtered functional principal components and $\left( \delta_{1},\ldots ,\delta_{s}\right)^{\prime }$ the scalar parameters associated to non-functional variables. By using a reduced set of $q_1,q_2,\ldots,q_R$ filtered functional principal components $\Gamma_{1(q_1)}^{\ast},\Gamma_{2(q_2)}^{\ast},\ldots,\Gamma_{R(q_R)}^{\ast},$ respectively, the model is then expressed as
  \begin{eqnarray*}
  	L&=& \mathbf{1}\alpha + \Gamma_{1(q_1)}^{\ast} \gamma_1^{\ast} + \Gamma_{2(q_2)}^{\ast} \gamma_2^{\ast} + \cdots + \Gamma_{R(q_R)}^{\ast} \gamma_R^{\ast} +U_1 \delta_1+\cdots+U_S \delta_S.
  \end{eqnarray*}
  
  Basis coefficients for each functional parameter are then obtained by formula (\ref{pclogitfun}) from their corresponding $\gamma^{\ast}$ parameters and the Intercept $\alpha^{\ast}$ by formula (\ref{Intercept}).
  
  The function of the \code{logitFD} package that allows fitting the filtered functional principal components logit regression model is \code{logitFD.fpc}. The performance of the function is the same as the \code{logitFD.pc} function.
  
  In order to illustrate the performance of the functions, let us again consider \code{StationsVar\$North} as binary response variable, \code{TempMonth} and \code{PrecMonth} as functional predictors, and as scalar predictor variables \code{StationsVar[,c("altitude","longitude")]}. We are going to consider the first 3 and 4 functional principal components of \code{TempMonth} and \code{PrecMonth}, respectively.
  
  Our fit is obtained as
  
  \begin{verbatim}
  	Fit2 <- logitFD.fpc(Response=StationsVars$North,FDobj=list(TempMonth.fd,PrecMonth.fd),
  	ncomp = c(3,4),nonFDvars = StationsVars[,c("altitude","longitude")])
  \end{verbatim}
  
  The output of this function is an \code{R} list with the same elements that were explained in the previous section. Next, the results of the fit are shown.
  
  \paragraph{\code{glm.fit} object of \code{Fit2}:} explained in the previous Section, its results can be seen next to the ones obtained for \code{Fit1} 
  
  {\scriptsize
  	\noindent
  	\begin{minipage}[t]{.49\textwidth}
  		\raggedleft
  		\begin{verbatim}
  			-----------------------------------------------------
  			summary(Fit1$glm.fit)
  			
  			Call:
  			glm(formula = design, family = binomial)
  			
  			Deviance Residuals: 
  			Min        1Q    Median        3Q       Max  
  			-1.77059  -0.01185   0.00000   0.01309   2.02115  
  			
  			Coefficients:
  			Estimate Std. Error z value Pr(>|z|)  
  			(Intercept) 15.10398    8.80373   1.716   0.0862 .
  			A.1         -1.94776    1.05278  -1.850   0.0643 .
  			A.2         -0.19686    0.58414  -0.337   0.7361  
  			A.3         -6.69297    3.49893  -1.913   0.0558 .
  			B.1          0.41633    0.78514   0.530   0.5959  
  			B.2          0.51503    6.42736   0.080   0.9361  
  			B.3         -3.11044    3.06542  -1.015   0.3103  
  			B.4         -2.44083    5.69108  -0.429   0.6680  
  			altitude    -0.02846    0.01576  -1.806   0.0709 .
  			longitude    1.40203    0.85922   1.632   0.1027  
  			---
  			(Dispersion parameter for binomial family taken to be 1)
  			
  			Null deviance: 100.857  on 72  degrees of freedom
  			Residual deviance:  14.785  on 63  degrees of freedom
  			AIC: 34.785
  			
  			Number of Fisher Scoring iterations: 15
  			------------------------------------------------------
  		\end{verbatim}
  	\end{minipage}
  	\hfill
  	\begin{minipage}[t]{.49\textwidth}
  		\raggedleft
  		\begin{verbatim}
  			------------------------------------------------------
  			summary(Fit2$glm.fit)
  			
  			Call:
  			glm(formula = design, family = binomial)
  			
  			Deviance Residuals: 
  			Min          1Q      Median          3Q         Max  
  			-2.671e-04  -2.100e-08   2.100e-08   2.100e-08   2.939e-04  
  			
  			Coefficients:
  			Estimate Std. Error z value Pr(>|z|)
  			(Intercept)    598.163  75918.975   0.008    0.994
  			A.1            -99.485  11468.867  -0.009    0.993
  			A.2            -13.281  10608.315  -0.001    0.999
  			A.3           -264.950  45230.675  -0.006    0.995
  			B.1             26.123   7055.585   0.004    0.997
  			B.2            174.667  31244.941   0.006    0.996
  			B.3            318.127  79802.859   0.004    0.997
  			B.4           -828.247 233825.008  -0.004    0.997
  			altitude        -1.251    142.820  -0.009    0.993
  			longitude       50.865   7090.131   0.007    0.994
  			---
  			(Dispersion parameter for binomial family taken to be 1)
  			
  			Null deviance: 1.0086e+02  on 72  degrees of freedom
  			Residual deviance: 2.2821e-07  on 63  degrees of freedom
  			AIC: 20
  			
  			Number of Fisher Scoring iterations: 25
  			------------------------------------------------------
  		\end{verbatim}
  	\end{minipage}
  }
  \label{glmsummary}
  
  \paragraph{Classification table of \code{Fit2}:} obtained as explained for \code{Fit1}, we can observe that the filtered functional principal components provide a better fit with CCR of 100\% in spite a less accurate estimation of parameters due to the high standard error of coefficients estimation. This fact was observed and stated in \cite{Aguilera2008151} for example. 
  
  \paragraph{\code{Intercept} object of \code{Fit2}:} provides the same result seen in the \code{Fit1} case.
  
\begin{figure}
  	\begin{center}
  		\begin{tabular}{ccc}
  			\includegraphics[width=0.33\textwidth]{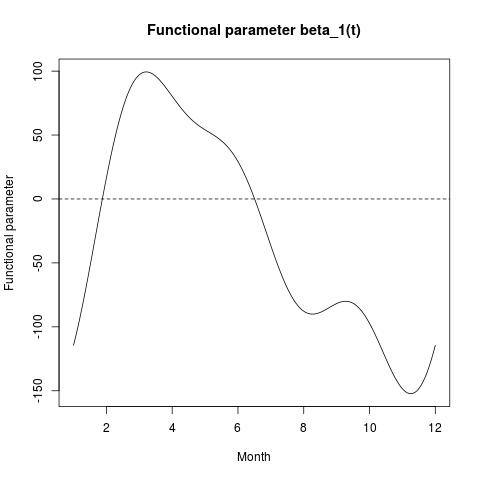} & \includegraphics[width=0.33\textwidth]{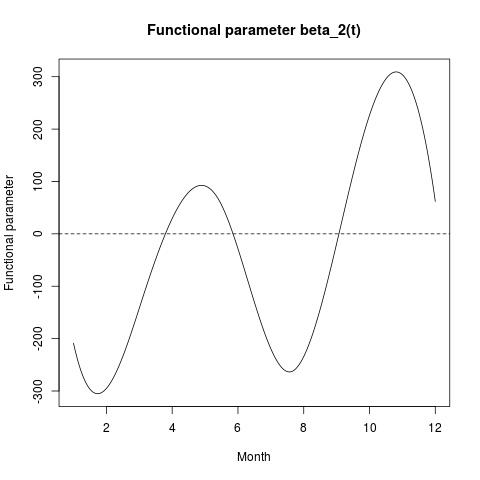} 
  			& \includegraphics[width=0.33\textwidth]{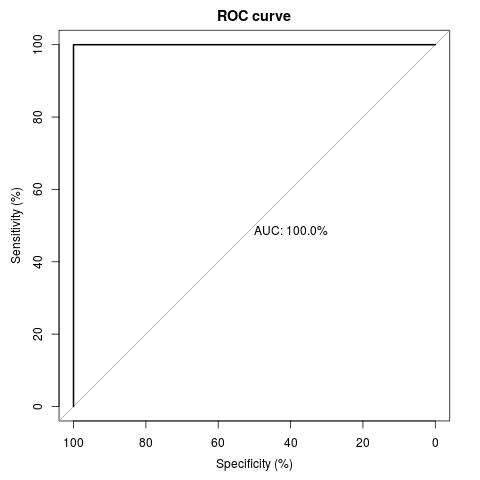}\\
  		\end{tabular}
  	\end{center}
  	\caption{Functional parameters and Roc Curve from Fit 2 by means of \code{logitFD.fpc()} function.}
  	\label{Fit2}
  \end{figure}

  \paragraph{\code{betalist} object of \code{Fit2}:} The functional parameters obtained by this fit can be seen in Figure \ref{Fit2}. We can observe the great similarity of the functional parameters form provided by the fit in terms of ordinary functional principal components and in terms of filtered functional principal components. The evaluation of these functions in the observed months-time, appears next with the ones obtained for \code{Fit1}:
  
  \noindent
  \begin{minipage}[t]{.49\textwidth}
  	\raggedleft
  	\begin{verbatim}
  		Fit1
  		--------------------------------
  		Months      Beta1      Beta2
  		1     Jan -2.6186728 -0.3861178
  		2     Feb  0.6541213 -0.2615488
  		3     Mar  2.6530440 -0.5074478
  		4     Apr  2.2007300  0.4053187
  		5     May  1.4871676  1.3946294
  		6     Jun  0.7918409  0.3944050
  		7     Jul -0.9034488 -1.8776571
  		8     Aug -2.1700722 -2.6123256
  		9     Sep -1.9520934 -1.1094367
  		10    Oct -2.2614869  0.5243271
  		11    Nov -3.4888495  0.9816193
  		12    Dec -2.6186728  1.1577407
  		--------------------------------
  	\end{verbatim}
  \end{minipage}
  \hfill
  \begin{minipage}[t]{.49\textwidth}
  	\raggedleft
  	\begin{verbatim}
  		Fit2
  		--------------------------------
  		Months      Beta1      Beta2
  		1     Jan -114.45858 -208.85221
  		2     Feb   16.52657 -295.08973
  		3     Mar   97.13723 -142.07106
  		4     Apr   80.32782   29.22866
  		5     May   54.09383   91.48928
  		6     Jun   29.28698  -28.35074
  		7     Jul  -36.52664 -219.37744
  		8     Aug  -87.77921 -234.26248
  		9     Sep  -81.81846  -18.66271
  		10    Oct  -97.27569  226.44071
  		11    Nov -148.43166  303.70498
  		12    Dec -114.45858   61.99109
  		--------------------------------
  	\end{verbatim}
  \end{minipage}
  \label{BasisEval}
  
  \noindent The code used for generating these results is \code{data.frame("Months" = names(monthLetters), "Beta1" = eval.fd(c(1:12), Fit1\$betalist[[1]]), "Beta2" = eval.fd(c(1:12), Fit1\$betalist[[2]]))} for the left-hand side and changing \code{Fit1} by \code{Fit2} for the right-hand side.  
  
  \paragraph{\code{PC.variance} object of \code{Fit2}:} it can be observed that there are several differences in the dynamic of variance accumulation between ordinary and filtered functional principal component analysis.
  
  \noindent
  \begin{minipage}[t]{.49\textwidth}
  	\raggedleft
  	\begin{verbatim}
  		----------------------------------
  		Fit1$PC.variance
  		[[1]]
  		Comp. % Prop.Var % Cum.Prop.Var
  		1   A.1       85.9           85.9
  		2   A.2       13.5           99.4
  		3   A.3        0.4           99.8
  		4   A.4        0.1           99.9
  		5   A.5        0.0           99.9
  		6   A.6        0.0           99.9
  		7   A.7        0.0           99.9
  		[[2]]
  		Comp. % Prop.Var % Cum.Prop.Var
  		1   B.1       98.2           98.2
  		2   B.2        0.9           99.1
  		3   B.3        0.6           99.7
  		4   B.4        0.3          100.0
  		5   B.5        0.1          100.1
  		6   B.6        0.0          100.1
  		7   B.7        0.0          100.1
  		8   B.8        0.0          100.1
  		----------------------------------
  	\end{verbatim}
  \end{minipage}
  \hfill
  \begin{minipage}[t]{.49\textwidth}
  	\raggedleft
  	\begin{verbatim}
  		----------------------------------
  		Fit2$PC.variance
  		[[1]]
  		Comp. % Prop.Var % Cum.Prop.Var
  		1   A.1     85.888         85.888
  		2   A.2     13.479         99.367
  		3   A.3      0.440         99.807
  		4   A.4      0.132         99.939
  		5   A.5      0.034         99.973
  		6   A.6      0.016         99.989
  		7   A.7      0.010         99.999
  		[[2]]
  		Comp. % Prop.Var % Cum.Prop.Var
  		1   B.1     99.070         99.070
  		2   B.2      0.536         99.606
  		3   B.3      0.311         99.917
  		4   B.4      0.049         99.966
  		5   B.5      0.031         99.997
  		6   B.6      0.002         99.999
  		7   B.7      0.000         99.999
  		8   B.8      0.000         99.999
  		----------------------------------
  	\end{verbatim}
  \end{minipage}
  \label{VarAcum}

  \paragraph{\code{ROC.curve} object of \code{Fit2}:} explained in the previous Section, the plot of the ROC curve appears Figure \ref{Fit2}. This graph and the area under the ROC curve (100\%) show an improvement of the prediction ability of the fit with filtered functional principal components in comparison with ordinary functional principal components.
  
  \section{Ordinary and filtered functional principal components included in the model according to their prediction ability (stepwise method)}
  
  \cite{Escabias04} proposed two alternative methods to include functional principal components in the logit model for both FPCA types: ordinary or filtered. On the one hand, functional principal components would be able to be included in the model in the order given by their explained variability. In that case the user should decide the number of functional principal components to include in the model for getting an accurate estimation of the functional parameter or for getting good prediction ability for the response. On the other hand, an automatic selection method of functional principal components could be used by a stepwise method. In this case the prediction ability of functional principal components would be the criterium to select the functional principal components and data would be the responsible of the model fit and prediction.
  
  \CRANpkg{logitFD} package contains two functions to fit the functional logit model after a stepwise selection procedure of functional principal components (ordinary and filtered) and nonfunctional variables. The fits obtained by these stepwise procedures are shown next. 
  
  \begin{verbatim}
  	Fit3 <- logitFD.pc.step(Response=StationsVars$North,FDobj=list(TempMonth.fd,PrecMonth.fd),
  	nonFDvars = StationsVars[,c("altitude","longitude")])
  	Fit4 <- logitFD.fpc.step(Response=StationsVars$North,FDobj=list(TempMonth.fd,PrecMonth.fd),
  	nonFDvars = StationsVars[,c("altitude","longitude")])
  \end{verbatim}
  
  Let us observe that for these functions it is not necessary to use a number of components parameter in the functions. We call \code{Fit3} for ordinary functional principal component analysis and \code{Fit4} for filtered functional principal component analysis. 
  
  The output of these function are \code{R} lists with the same elements as the ones seen in \code{Fit1} and \code{Fit2}. We only show and explain here some of the results.
  
  \paragraph{\code{glm.fit} objects of \code{Fit3} and \code{Fit4}:} We can observe from these results that stepwise method selected three functional principal components for Temperature and only one for Precipitation. Regarding scalar predictors, the method selected the altitude variable. Note that stepwise selection included the same components for both approaches, although the values of their parameters and standard errors are different. The classification ability of these fits is 100\% of correct classification rate and can be obtained by using the same code shown fof \code{Fit1} and \code{Fit2}.
  
  {\scriptsize
  	\noindent
  	\begin{minipage}[t]{.49\textwidth}
  		\raggedleft
  		\begin{verbatim}
  			-------------------------------------------------------
  			summary(Fit3$glm.fit)
  			
  			Call:
  			glm(formula = Response ~ A.1 + altitude + A.7 + A.3 + B.5,
  			family = binomial, data = design)
  			
  			Deviance Residuals: 
  			Min          1Q      Median          3Q         Max  
  			-3.677e-04  -2.000e-08   2.000e-08   2.000e-08   2.960e-04  
  			
  			Coefficients:
  			Estimate Std. Error z value Pr(>|z|)
  			(Intercept)    936.784  71065.432   0.013    0.989
  			A.1           -223.554  16658.601  -0.013    0.989
  			altitude        -2.543    191.207  -0.013    0.989
  			A.7           4016.721 300525.378   0.013    0.989
  			A.3           -972.450  73148.168  -0.013    0.989
  			B.5            308.717  23326.153   0.013    0.989
  			
  			(Dispersion parameter for binomial family taken to be 1)
  			
  			Null deviance: 1.0086e+02  on 72  degrees of freedom
  			Residual deviance: 4.7820e-07  on 67  degrees of freedom
  			AIC: 12
  			
  			Number of Fisher Scoring iterations: 25
  			------------------------------------------------------
  		\end{verbatim}
  	\end{minipage}
  	\hfill
  	\begin{minipage}[t]{.49\textwidth}
  		\raggedleft
  		\begin{verbatim}
  			------------------------------------------------------
  			summary(Fit4$glm.fit)
  			
  			Call:
  			glm(formula = Response ~ A.1 + altitude + A.7 + A.3 + B.5,
  			family = binomial, data = design)
  			
  			Deviance Residuals: 
  			Min          1Q      Median          3Q         Max  
  			-6.974e-04  -2.000e-08   2.000e-08   2.000e-08   5.753e-04  
  			
  			Coefficients:
  			Estimate Std. Error z value Pr(>|z|)
  			(Intercept)  1.938e+03  3.312e+05   0.006    0.995
  			A.1         -3.899e+02  3.754e+04  -0.010    0.992
  			altitude    -4.731e+00  6.218e+02  -0.008    0.994
  			A.7          6.724e+03  1.132e+06   0.006    0.995
  			A.3         -1.557e+03  8.121e+04  -0.019    0.985
  			B.5          6.409e+02  6.659e+04   0.010    0.992
  			
  			(Dispersion parameter for binomial family taken to be 1)
  			
  			Null deviance: 1.0086e+02  on 72  degrees of freedom
  			Residual deviance: 1.1402e-06  on 67  degrees of freedom
  			AIC: 12
  			
  			Number of Fisher Scoring iterations: 25
  			------------------------------------------------------
  		\end{verbatim}
  	\end{minipage}
  }
  
\begin{figure}
  	\begin{center}
  		\begin{tabular}{ccc}
  			\includegraphics[width=0.33\textwidth]{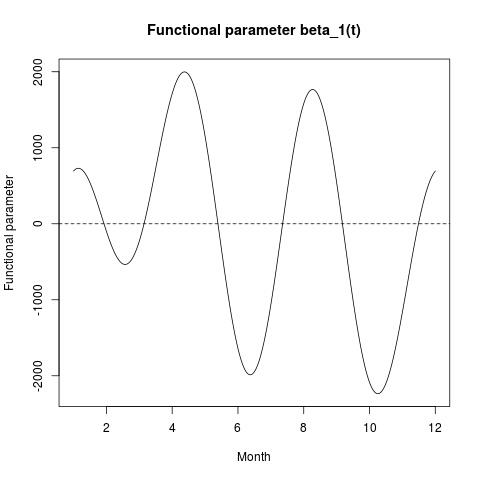} & \includegraphics[width=0.33\textwidth]{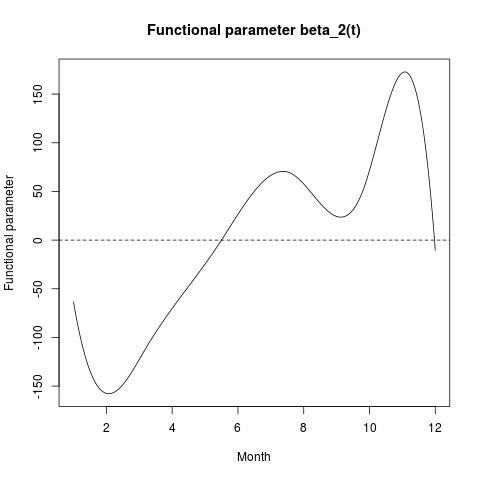} & \includegraphics[width=0.33\textwidth]{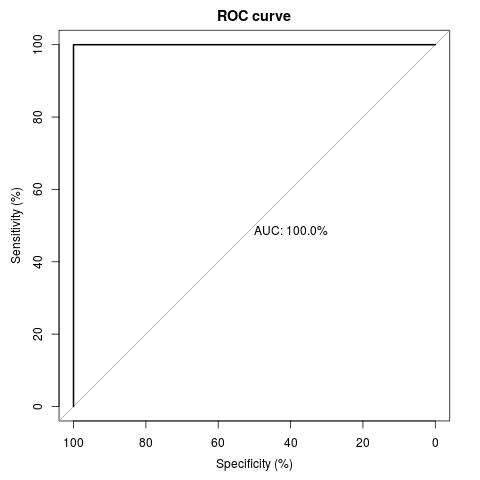} \\
  			\multicolumn{3}{c}{Ordinary FCPA and stepwise order}\\
  			\includegraphics[width=0.33\textwidth]{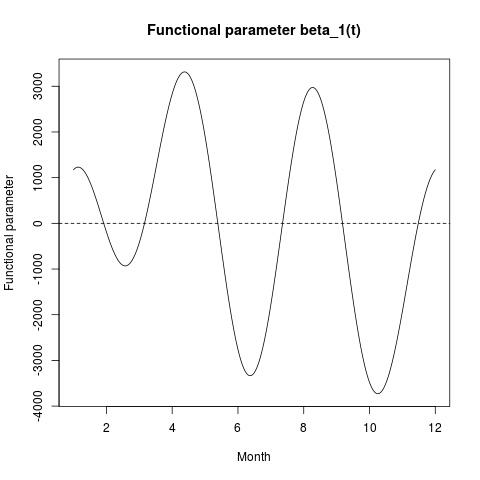} & \includegraphics[width=0.33\textwidth]{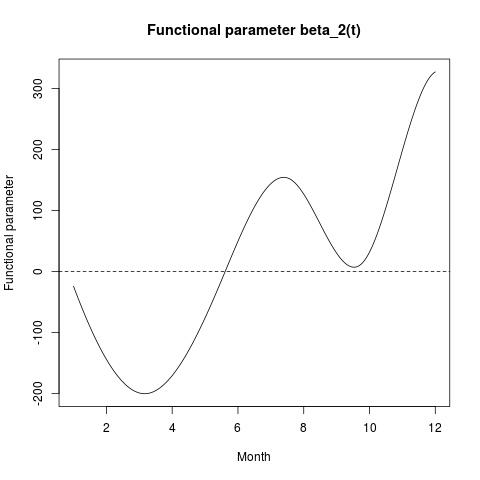} & \includegraphics[width=0.33\textwidth]{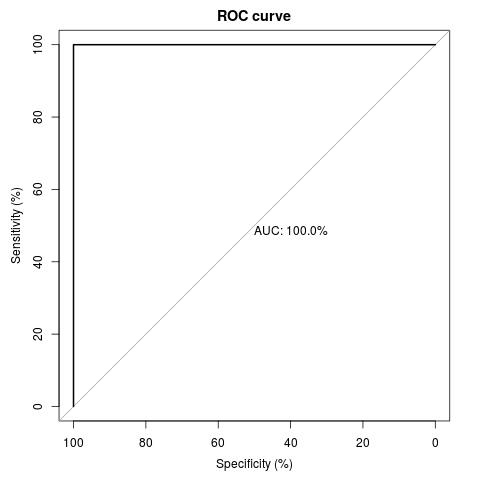}\\
  			\multicolumn{3}{c}{Filtered FCPA and stepwise order}\\
  		\end{tabular}
  	\end{center}
  	\caption{Functional parameters and Roc Curve from Fits 3 and 4 by means of \code{logitFD.pc.step()} and \code{logitFD.fpc.step()} functions respectively.}
  	\label{Fit34}
  \end{figure}

  \paragraph{\code{betalist} objects of \code{Fit3} and \code{Fit4}:} The graphs of estimated functional parameters are shown in Figure \ref{Fit34}. It can be seen the similarity in the forms of the functional parameters, in spite of the evaluation values are different as can be seen next:
  \begin{center}
  	\begin{verbatim}
  		-----------------------------------------------------
  		Months  Fit3.Beta1 Fit3.Beta2 Fit4.Beta1 Fit4.Beta2 
  		1     Jan   693.10831  -63.47274  1172.6949  -24.28034
  		2     Feb   -98.76707 -157.45404  -186.0346 -144.18907
  		3     Mar  -229.11217 -122.61836  -423.5395 -199.28285
  		4     Apr  1711.91853  -70.18329  2831.9461 -170.88030
  		5     May  1152.29655  -24.62758  1904.3853  -76.25583
  		6     Jun -1640.56224   26.48668 -2761.1827   49.39343
  		7     Jul -1066.07148   66.45460 -1780.0473  143.75270
  		8     Aug  1580.73125   57.60667  2663.1874  126.59096
  		9     Sep   543.54551   24.31157   921.5509   29.89969
  		10    Oct -2086.28319   71.63066 -3481.0951   31.57623
  		11    Nov -1163.57269  171.95001 -1925.9015  198.33040
  		12    Dec   693.10831  -10.48963  1172.6949  326.95671
  		-----------------------------------------------------
  	\end{verbatim}
  \end{center}
  \noindent The code used for thsese evaluations was \code{data.frame("Months" = names(monthLetters), "Fit3.Beta1" = eval.fd(c(1:12), Fit3\$betalist[[1]]), 
  	"Fit3.Beta2" = eval.fd(c(1:12), Fit3\$betalist[[2]]), "Fit4.Beta1" = eval.fd(c(1:12), Fit4\$betalist[[1]]), "Fit4.Beta2" = eval.fd(c(1:12), Fit4\$betalist[[2]]))}

  \paragraph{\code{PC.variance} objects of \code{Fit3} and \code{Fit4}:} The objects of variance accumulation of the different functional principal components analysis do not change from the ones shown in previous sections. We do not show them here, the reader can check these equalities through the objects \code{Fit3\$PC.variance} and \code{Fit4\$PC.variance}. 
  
  \paragraph{\code{ROC.curve}  objects of \code{Fit3} and \code{Fit4}:} Roc objects with Roc areas provide an area under the roc curve of 100\% in each case. The plot of the ROC curves showing the good performance of the fits can be seen in Figure \ref{Fit34}.  
  
  \section{Conclusions}
  
  In this work the functions of the \CRANpkg{logitFD} package have been shown for fitting an extended functional principal components logistic regression model. The package provides two alternative solutions (ordinary and filtered FPCA) for the multicollinearity problem that arises when the functional predictors and the parameter functions are assumed to belong to the same finite dimensional space generated by a basis of functions. The dimension of the basis can be different in each functional variable in the model. Likewise, for each of the proposed solutions, two ways of choosing the functional principal components are provided: on the one hand, the users must manually choice the adequate number of components to be included in the model in order of variability, i.e., the first $q$ principal components that overcome a certain variability percentage; on the other hand in the automatic order provided by the stepwise method, that is, according to predictive ability of principal components and non-functional variables.
  
  The illustration of the use of the package's functionalities has been carried out using a set of functional and non-functional data, included in the \CRANpkg{fda.usc} package. In particular, weather functional variables observed in 73 Spanish weather stations, such as the mean monthly evolution of temperatures and rainfall, and non-functional as the spatial location of the weather stations in the Spanish territory are considered throughout the current manuscript.
  
  The conclusions we have reached after the fits can be summarized in that the variables that best describe the North-South location of the meteorological stations are the mean monthly precipitation and temperature (through their first, third and seventh principal components for temperature and fifth for rainfall) and the own altitude of the weather stations. All the models provide good predictive ability, with the solutions based on ordinary and filtering FPCA by stepwise selection being  the best (100 \% CCR) due to their balance between reduced dimension and predictive ability. Likewise, the filtered FPCA solution including the components in order of variability provides results equally good to the previous ones but with more variables. The ordinary FPCA-based solution including the components in order of variability provides results similar to those previously described.
  
  As was stated in the Introduction section, the \code{fregre.glm} function of the \CRANpkg{fda.usc} package aim to achieve the same goal as the functions included in \CRANpkg{logitFD} package, but through different point of view: \code{fregre.glm} use a discrete based methodology of functional data and \CRANpkg{logitFD} functions use a purely functional approach using fd objects from the \CRANpkg{fda} package. This approach makes the functional models of scalar response to suffer of multicollinearity problems with the inaccurate estimation of the functional parameters as a consequence (see \cite{Escabias04}). Two solutions based on functional PCA are implemented in \CRANpkg{logitFD} package: (1) classic functional PCA and (2) filtered functional PCA. Each of PCA methods have been revealed to be useful in a different aspect: the first allow a lower estimation error of the basic coefficients of the functional parameters, while the second allow a lower estimation error of the proper curve, in terms of mean integrated quadratic error (see \cite{Escabias04}). Moreover the literature has also shown for methods involving principal components, that sometimes principal components with low variability explanation can be good predictors of the response, so a stepwise selection method of functional principal components has been included. So the main difference among \CRANpkg{logitFD} functions and \code{fregre.glm} is that all the mentioned issues are addressed in the \CRANpkg{logitFD} package and solved in a fast and transparent way and they are not taken into account in \code{fregre.glm}. Finally it is important to point out that the output of the functional elements of the \CRANpkg{logitFD} functions (as the functional parameters) are also \code{fd} objects and therefore all the functions of the \CRANpkg{fda} package could be used with them for plotting, evaluating, etc.
  
  In short, logitFD package provides its users with the possibilities to deal with the functional logit regression model from basis expansion methodology of sample curves and solving in a fast and transparent way, the problems that arise through functional principal component analysis. In our opinion, if we wanted to solve the same problems by using alternative R functions with similar goal, it would be necessary give many steps that would make the process to be highly tedious. For this reason, and due to logit regression is highly considered in real problems, we think that the current manuscript can be very interesting for the readers given that they could use it as reference manual in their analysis
  
\begin{figure}
  	\begin{center}
  		\begin{tabular}{ccc}
  			\includegraphics[width=1.0\textwidth]{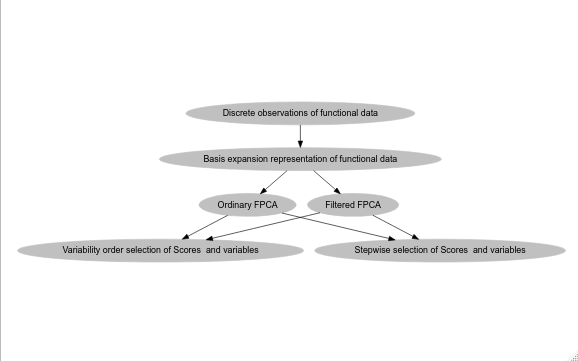}
  		\end{tabular}
  	\end{center}
  	\caption{Steps for Functional Principal Componets Logit Regression fit in its different situations considered in \CRANpkg{logitFD} package.}
  	\label{Diag}
  \end{figure}

  Figure \ref{Diag} give a schematic diagram that summarize the steps of the  methodology followed along the paper
  
  \section{Acknowledgment}
  
  This paper is partially supported by the project FQM-307 of the Government of Andalusia (Spain) and  by the project PID2020-113961GB-I00 of the Spanish Ministry of Science and Innovation (also supported by the FEDER programme). They also acknowledge the financial support of the Consejería de Conocimiento, Investigación y Universidad, Junta de Andalucía (Spain) and the FEDER programme for project A-FQM-66-UGR20. Additionally, the authors acknowledge financial support by the IMAG–María de Maeztu grant CEX2020-001105-M/AEI/10.13039/501100011033.
  
  \bibliography{Escabias-Aguilera-Acal}
  
  \address{Manuel Escabias\\
  	Department of Statistics and Operation Research, University of Granada\\
  	Facultad de Farmacia, Campus de Cartuja. 18071 Granada\\
  	Spain\\
  	ORCiD: 0000-0002-1653-9022\\
  	\email{escabias@ugr.es}}
  
  \address{Ana M. Aguilera\\
  	Department of Statistics and Operation Research, University of Granada\\
  	Facultad de Ciencias, Campus Fuentenueva. 18071 Granada\\
  	Spain\\
  	ORCiD: 0000-0003-2425-6716\\
  	\email{aaguiler@ugr.es}}
  
  \address{Christian Acal\\
  	Department of Statistics and Operation Research, University of Granada\\
  	Facultad de Ciencias, Campus Fuentenueva. 18071 Granada\\
  	Spain\\
  	ORCiD: 0000-0002-2636-5396\\
  	\email{chracal@ugr.es}}
  
\end{article}

\end{document}